\documentclass[usegraphicx,usenatbib]{mn2e}

\usepackage{fixltx2e}
\usepackage{mathptmx}

\newcommand{\Mpc}{\rm\thinspace Mpc}

\newcommand{\km}{\rm\thinspace km}

\newcommand{\cm}{\rm\thinspace cm}
\newcommand{\s}{\rm\thinspace s}

\newcommand{\Msun}{\hbox{$\rm\thinspace M_{\odot}$}}

\newcommand{\keV}{\rm\thinspace keV}

\newcommand{\erg}{\rm\thinspace erg}

\newcommand{\ergpcmsqps}{\hbox{$\erg\cm^{-2}\s^{-1}\,$}}

\newcommand{\ergps}{\hbox{$\erg\s^{-1}\,$}}

\newcommand{\kmps}{\hbox{$\km\s^{-1}\,$}}

\newcommand{\kmpspMpc}{\hbox{$\kmps\Mpc^{-1}$}}

\newcommand{\pcmsq}{\hbox{$\cm^{-2}\,$}}

\begin{document}

\title[A ghost in the CDFN]{The extended X-ray emission around
  HDF\,130 at z=1.99: an inverse Compton ghost of a giant radio source
  in the Chandra Deep Field North} \author[Fabian et al
]{A.C. Fabian$^1\thanks{E-mail: acf@ast.cam.ac.uk}$,
  S. Chapman$^1$, C.M. Casey$^1$, F. Bauer$^2$ and K.M. Blundell$^3$\\
  1. Institute of Astronomy, Madingley Road, Cambridge. CB3 0HA\\
  2. Columbia Astrophysics Laboratory, 550 W. 120th St.,
  Columbia University, New York, NY 10027, USA\\
  3.University of Oxford, Department of Physics, Keble Road, Oxford
  OX1 3RH }
\maketitle
  
\begin{abstract}
  One of the six extended X-ray sources found in the Chandra Deep
  Field North is centred on HDF\,130, which has recently been shown to
  be a massive galaxy at $z=1.99$ with a compact radio nucleus. The
  X-ray source has a roughly double-lobed structure with each lobe
  about 41 arcsec long, or 345~kpc at the redshift of HDF\,130. We
  have analyzed the 2~Ms X-ray image and spectrum of the source and
  find that it is well fit by a power-law continuum of photon index
  2.65 and has a 2--10~keV luminosity of $5.4\times 10^{43}\ergps$ (if
  at $z=1.99$). Any further extended emission within a radius of
  60~arcsec has a luminosity less than half this value, which is
  contrary to what is expected from a cluster of galaxies. The source
  is best explained as an inverse Compton ghost of a giant radio
  source, which is no longer being powered, and for which Compton
  losses have downgraded the energetic electrons, $\gamma> 10^4,$
  required for high-frequency radio emission.  The lower energy
  electrons, $\gamma\sim 1000,$ produce X-rays by inverse Compton
  scattering on the Cosmic Microwave Background. Depending on the
  magnetic field strength, some low frequency radio emission may
  remain. Further inverse Compton ghosts may exist in the Chandra deep
  fields and beyond.
\end{abstract}


\section{Introduction}

The deepest published X-ray image of the Sky is the 2~Ms Chandra Deep
Field North (CDFN: Alexander et al 2003). Bauer et al (2002) studied
the 6 extended sources found in the first Ms exposure of this field
and inferred that they were all due to clusters and groups. Casey et
al (2008) have recently found that object HDF\,130, at the centre of
one of these extended sources (Source 2, CDFN\,J123620.0+621554, of
Bauer et al 2002), is a massive elliptical galaxy at redshift
$z=1.99$, and not a starburst galaxy as earlier supposed. HST/NICMOS
and Spitzer/IRAC imaging have revealed that the object has a massive
($\sim 3\times 10^{11}\Msun$) old stellar population with a large
diameter of about 8~kpc. It is intrinsically weak at submillimetre
wavelengths and has a compact radio source, indicative of a central
Active Galactic Nucleus (AGN).  We investigate here the possibility
that the double-lobed structure of the extended X-ray source is not
then due to a surrounding merging cluster, as suggested by Bauer et al
(2002), but instead to inverse Compton (IC) emission from a past
outburst of the galaxy. The object would in the past have appeared as
a giant radio galaxy, but inverse Compton losses have downshifted the
high energy electrons responsible for radio emission so what remains
is X-ray IC emission produced by Lorentz factor $\gamma\sim 1000$
electrons scattering on the Cosmic Microwave Background (CMB).

Extensive IC X-ray emission from large radio galaxies has been
detected above redshift $z=1$, such as 3C\,294 ($z=1.786$; Fabian et
al 2003), 6C\,0905+39 ($z=1.833$, Blundell et al 2006; Erlund et al
2008) and 4C\,23.56 ($z=2.48$; Johnson et al 2007). The flux of the
emission depends on the energy density of the target photons, which in
the case of the CMB rises as $(1+z)^4$ so cancelling out the dimming
expected from increased distance (Felten \& Rees 1969; Schwartz
2002). In the case of 3C\,294 there are distinct large patches of
X-ray emission separate from the observed radio source (Erlund et al
2006) and in 6C\,0905+39 most of the X-ray emission lies in between
the hotspots where no radio emission is seen (Erlund et al 2008). The
lifetime of the electrons in the sources scales as $1/\gamma$ due to
radiative losses.  Lorentz factors of $\gamma\sim 1000$ are required
to upscatter the CMB photons and $\gamma\sim 10^{4}$ (depending on the
magnetic field strength) to generate GHz synchrotron radiation in the
radio band. Consequently when the AGN switches off and no further
electron acceleration takes place, the IC X-ray emission can last ten
or more times longer than the high-frequency radio source, as observed
with current instruments.

\begin{figure*}
  \includegraphics[width=0.7\textwidth]{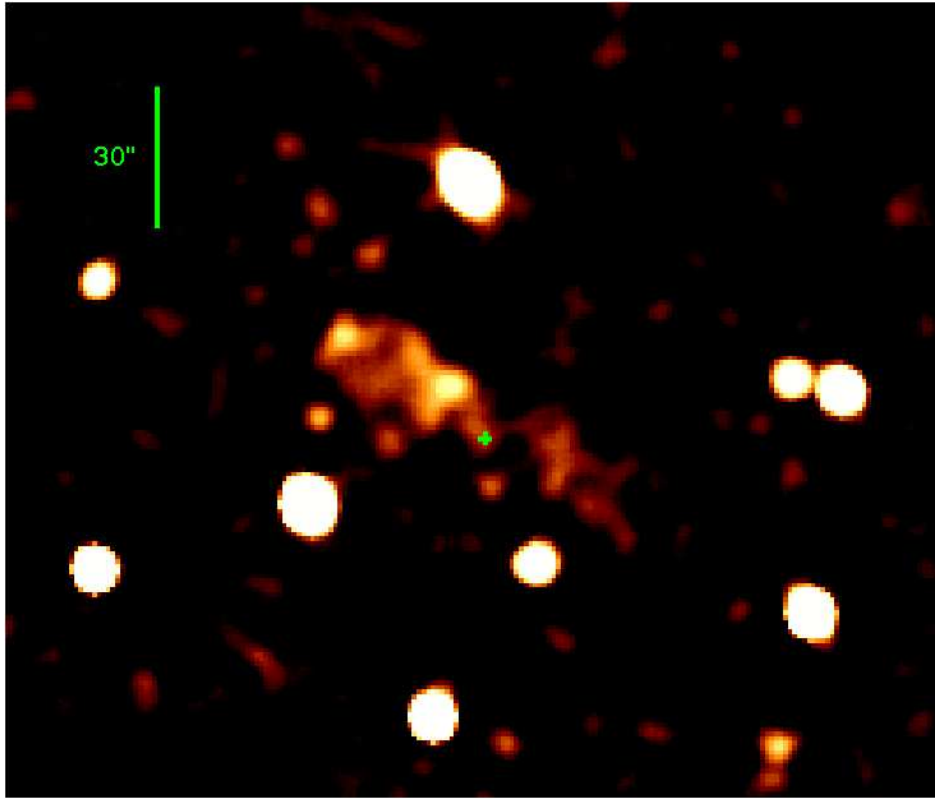}
  \caption{X-ray image of the source in the 0.3--3~keV band. The data
    have been binned into 1~arcsec pixels and gaussian smoothed by 5
    pixels. 20 arcsec is indicated by the bar on the left and the
    position of the massive galaxy HDF\,130 hosting the (unresolved)
    radio source of Casey et al (2008) is marked by a cross. }
\end{figure*}

The Compton lifetime of $\gamma\sim 1000$ electrons at $z\sim 2$ is
about 30~Myr, so the X-ray ghost of a dead giant radio galaxy at that
redshift may be detectable as an extended X-ray source for that
time. Depending on the ratio of the radio to Compton lifetimes, and on
adiabatic expansion losses, such X-ray sources could be common.
Celotti \& Fabian (2004) considered the relative numbers of radio
galaxies and clusters as a function of redshift and showed that such
extended ghosts could well outnumber low luminosity clusters at
moderate to high redshifts. They predicted that such objects could
account for some of the extended objects in the Chandra deep fields.

Here we investigate whether Source 2 in the extended CDFN source list
of Bauer et al (2002) can be such a ghost of a radio source, given its
double-lobed structure with a massive galaxy hosting a compact radio
source at the centre. The case stands on the morphology and spectrum
of the X-ray source and the coincidence with the massive
galaxy. Fortunately the X-ray dataset has been doubled (from 1--2\,Ms
exposure) since the work of Bauer et al (2002).

The Hubble Deep Field region has been observed at radio frequencies
(1.4~GHz) at the VLA (Richards et al 1999; Biggs \& Ivison 2006) and
Westerbork (Garrett et al 2000). Two faint VLA radio sources,
unresolved in the 1.5 arcsec beam, appear at the position of the
extended X-ray emission\footnote{One is HDF\,130, which is unresolved
  ($<65$\,mas) by MERLIN (Casey et al 2008); the other, 10~arcsec to
  the NNW and coincident with an X-ray emission peak (Alexander et al
  2003), is a submillimetre galaxy at the same redshift as HDF130
  (Chapman et al 2008, submitted).}, with a flux totalling
$\sim400\mu{\rm Jy}$. This compares with a single unresolved WSRT
source of $540\mu{\rm Jy}$ at the same wavelength of 21~cm (see the
map in Garrett et al 2000 which shows contours of the Westerbork
source and marks the positions of the VLA sources). The much larger
beam of the Westerbork array ($\sim 15$~arcsec; the source size is
quoted as $<21$~arcsec) is more sensitive to extended radio flux which
can explain the difference between the measured fluxes. In other
words, the data appear consistent with the presence of an extended
1.4~GHz radio source of about $\sim150\mu{\rm Jy}$ at the position of
the extended X-ray source.  For the present, we treat it as an upper
limit.

We assume $H_0 = 70 \kmpspMpc$, which translates into a scale of 8.5~kpc
per arcsec at $z=1.99$.

\section{Data analysis}

The X-ray image (Fig.~1) shows an extended source which is roughly
linear in shape from NE to SW (aspect ratio of about 4 to 1) and with
a minimum in the middle. The point-like radio source lies at the
centre of the structure close to the minimum. Each lobe is 40.7~arcsec
(345~kpc) long, so the whole X-ray source is 690~kpc across. This is
not exceptional for a giant radio galaxy (e.g. Mullin et al 2008).

Thus on a purely morphological basis the source resembles what a ghost
radio source should look like and is unlike a relaxed cluster of
galaxies. Such a massive galaxy is plausibly in a cluster of some form
and it would not be surprising for there to be some extended hot gas
associated with it. An important test is how extended the X-ray
emission appears; a cluster atmosphere should be very extended and
appear roughly circular and not the limear shape we observe here.

The 2~Ms CDF-N data (see Alexander et al. 2003 for a full description
of the data) were re-reduced following a procedure identical to that
described by Luo et al. (2008) for the 2~Ms CDF-S data analysis. X-ray
spectra of our source were then extracted from individual reduced and
cleaned level 2 event files using routines and recipes from the ACIS
Extract software package (AE; Broos et al. 2002), which automates and
augments the basic reduction method within CIAO (v4.0). We adopted two
formal extraction regions centred on HDF\,130, an elliptical region
that encompassed only the high-surface brightness emission
(45\arcsec$\times$17\arcsec, $\theta=330^{\circ}$ i.e. major axis NE)
and a circular region equivalent to 500~kpc in radius at $z=1.99$ (60
arcsec radius).  Background spectra were extracted from an annulus
between 60 arcsec and 100 arcsec centred on our source. In all
regions, point sources from the catalog of Alexander et al. (2003)
were excluded using circular regions a factor of two larger than the
99 per cent encircled energy radius at that position
($\approx$4--10\arcsec). The individual spectra and associated
calibration products were then merged, and the combined background
spectrum was scaled to that of our source based on the mean ratio of
the effective areas. We note that this effective area ratio is nearly
constant over the entire energy range, with only a $\approx$3--4 per
cent variation from end-to-end. The spectra were grouped at a
signal-to-noise ratio of 5 and then fit with XSPEC (Version 12.4.0;
Arnaud 1996).

The spectra from the smaller elliptical and the larger circular
regions have both been fitted with power-law and thermal (Mekal)
models, absorbed by the Galactic column density of $N_{\rm H}=2.93
\times 10^{20}\pcmsq$. The observed spectra are similar (Fig.~2) and
show no significant flux difference. The limit on any further extended
emission from a hot medium in any surrounding cluster is a rest-frame
2--10~keV luminosity of $\sim 2 \times10^{43}\ergps$ (a temperature of
3~keV was assumed). For the elliptical region covering the obvious
emission in Fig.~1 we find similar values of $\chi^2=107$ for 112
degrees of freedom for both the power-law and thermal models. The
fitted photon index for the former $\Gamma=2.65\pm0.5$ and the
temperature for the latter $kT=2.9^{+1.9}_{-1.0}\keV$, where the
uncertainties are quoted at 90 per cent confidence levels. The
0.5--5~keV flux of the source is $2.4\times 10^{-15}\ergpcmsqps$
(i.e. 0.6~nJy at 1~keV).

\begin{figure}
  \centering
  \includegraphics[width=0.65\columnwidth,angle=-90]{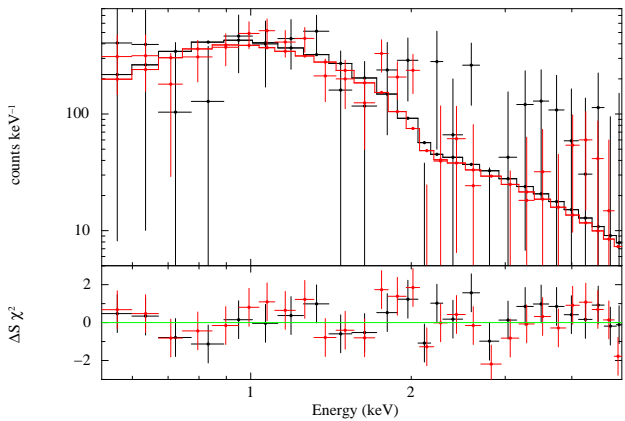}
  \caption{Spectra of the source; the black points are from the
    elliptical region immediately surrounding the extended emission
    seen in Fig.~1, the red points are from a larger circular region
    of radius 60~arcsec.}
\end{figure}

After correction for Galactic absorption, the 2--10~keV rest frame
luminosity of the source is $5.4\pm0.9\times 10^{43}\ergps$. This is
about one half of that of the giant powerful radio galaxy 6C\,0905+39
($1.5\times 10^{44}\ergps$; Erlund et al 2008), so the size and
luminosity of the extended CDFN X-ray source are not
exceptional. Although the spectrum does not distinguish power-law IC
from thermal emission, the linear shape of the object and lack of more
extensive emission strongly support a non-thermal origin; indeed there
is no known evidence of any thermal material within FRII lobes. The
photon index of the emission around HDF\,130 is steeper
($\Gamma\sim2.65$) than that associated with 6C\,0905+39
($\Gamma\sim1.8$) but only marginally so when uncertainties are
considered. This may indicate that spectral ageing is significant for
the electrons in the HDF\,130 source. The highest energy electrons
lose their energy at a faster rate than lower energy ones, so leading
to a steepening of the spectrum. It seems plausible that we are
observing HDF130 at a later stage in its lifecycle than we observe
6C0905+39.

\section{Discussion}
We have shown that Extended Source 2 from the CDFN (Bauer et al 2002)
is plausibly a ghost X-ray source with X-ray bright lobes. The total
electron energy associated with the emission is considerable at about
$7\times 10^{59}\erg$, using expression (6) in Erlund et al (2006)
with the electron distribution extending from $\gamma=10^3 - 10^5$ and
the above $2-10$~keV luminosity. This estimate can increase
considerably if the electron distribution is broader (particularly
extending below $\gamma=10^3$) and if energetic protons accompany the
electrons. At the present time we have neither direct or indirect
means to detect the protons, nor do we have any reason to invoke
particles with Lorentz factors below $10^3$ (those that are madnated
by the ICCMB emission). If the steep photon index is due to spectral
ageing then a turnover in the electron distribution is expected
there. The total energy of the source will have been larger in the
past if the radio lobes were over-pressured with respect to the
environment they were expanding into.  Also it is expected that
expansion losses will be important if the source is not confined in
some way. We presume this source does lie in a poor cluster or group
which has a hot medium with a low X-ray luminosity into which the
lobes will be doing work as they expand.

The expected radio flux from the source relies on several
assumptions. If we first assume that the magnetic field in the lobes
is in energy equipartition with the relativistic electrons estimated
above then we find a field of $2\mu$G. (The emission region has been
approximated as a cylinder 690~kpc long and 170~kpc in diameter.) This
means that the Lorentz factor for electrons giving 1.4~GHz synchrotron
radio emission is $\gamma\approx1.5\times10^4$. The synchrotron flux
can then be predicted from the inverse Compton flux using, for
example, expressions (4.53) and (4.54) of Tucker (1975). Assuming a
power-law electron distribution for $\gamma=10^3 - 10^5$, we find a
1.4~Ghz synchrotron flux of $\sim14\mu$Jy ($\sim 145\mu$Jy at 300~MHz)
for a photon index $\Gamma=2.6$ and $570\mu$Jy for $\Gamma=2$. Only
for a steep spectral index (consistent with an ageing electron
population) are the present 1.4~Ghz radio observations consistent with
this prediction.  The synchrotron flux $F_{\rm s}$ varies with the
total energy as $F_{\rm s}\propto E^{\Gamma/2}$, so a broader energy
distribution increases the predicted flux, thus strengthening this
conclusion. For example, if $\Gamma=2.6$ and the electron spectrum
turns over below $\gamma\sim 500$ then the predicted 1.4\,GHz
synchrotron flus is just consistent with the flux density of
$140\mu$Jy hinted at by the discrepancy between the Westerbork and VLA
images.

If the powerful jets have switched off and electron
acceleration ended within the hotspots then there will not be a simple
power-law electron distribution from $\gamma=10^3 - 10^4$, but a steep
turnover above the energy where the inverse Compton cooling time
equals the time $\Delta t$ since the acceleration stopped;
\begin{equation}
\gamma=1000\left({\Delta t}\over{30\,{\rm Myr}}\right)^{-1}.
\end{equation} 
This could mean that the radio source is only detectable at low radio
frequencies (few 100~MHz) and thus, for example, with the giant
metrewave radio telescope (GMRT) or in the future with Low Frequency
Array (LOFAR) and similar instruments.

Bauer et al (2002) note that there may be about 150 extended X-ray
sources per square degree. If even 10 per cent of them are similar to
the HDF\,130 source then they have a space density of roughly
$10^{-6}$~Mpc$^{-3}$ ($3\times 10^{-8}$~Mpc$^{-3}$ in comoving units),
close to the estimate of Celotti \& Fabian (2004) for sources of
luminosity $\sim 10^{44}\ergps$. The Compton cooling time at $z=2$ is
about 30~Myr, whereas the time from $z=2.5$ to 1.5 is about
1.7~Gyr. Therefore the space density of objects giving rise to ghosts
(assuming they each only undergo one jet-active phase) is 50 times the
above value or $\sim10^{-6}$~Mpc$^{-3}$ in comoving units, comparable
to the number density of massive galaxies ($M_{\rm K}<25$; Cole et al
2001). It is even higher if expansion losses dominate, as expected.

A direct demonstration of our IC explanation for the extended X-ray
emission around HDF130 would be the clear detection and mapping of
the radio source itself at low radio frequencies (e.g. 100~MHz), where
the Compton losses are less and the emitting electrons closer in
energy to the X-ray ones.

We conclude that the X-ray Sky may be littered with faint inverse
Compton ghosts.

\section*{Acknowledgements}
We acknowledge financial support from the Royal Society (ACF and KMB)
and {\it Chandra} grant SP8-9003B (FEB). CMC thanks the
Gates-Cambridge Trust.


\clearpage
\end{document}